\documentstyle[aps,12pt]{revtex}
\begin{document}
\title{Exact time evolution in harmonic quantum Brownian motion}
\author{Fabi\'an H. Gaioli and Edgardo T. Garcia-Alvarez}
\address{Instituto de Astronom\'\i a y F\'\i sica del Espacio, \\
C.C. 67, Suc. 28, 1428 Buenos Aires, Argentina\\
Departamento de F\'\i sica, Facultad de Ciencias Exactas y Naturales,\\
Universidad de Buenos Aires, 1428 Buenos Aires, Argentina}
\maketitle

\begin{abstract}
We consider a particular (exactly soluble) model of the one discussed in a
previous work. We show numerical results for the time evolution of the main
dynamical quantities and compare them with analytical results.
\end{abstract}

\bigskip\ Pacs: 05.40.+j


\bigskip\ Send proof to: Fabi\'an H. Gaioli,

Instituto de Astronom\'\i a y F\'\i sica del Espacio,

C.C. 67, Suc. 28, 1428 Buenos Aires, Argentina

e-mail: gaioli@iafe.uba.ar

fax: (54-1) 786-8114

TE: (54-1) 781-6755

\newpage\ 

\section{Introduction}

In Ref. \cite{primo} (hereafter referred as I) we have obtained a
generalized and exact form of the master and Langevin equations. In this
work we consider a particular soluble model and numerical results related to
such equations. We show that from the microscopic quantum-mechanical laws a
sort of $`$`irreversible'' behavior emerges under the following conditions:
a privileged initial condition (no correlation between the Brownian
oscillator and the bath), a relevance criterion arising from the form of the
Hamiltonian, which distinguishes the Brownian particle from the bath, and a
natural extra-dynamical hypothesis in order to interpret the exact dynamical
evolution. That is, to consider such a process during a time scale of
observation smaller than the recurrence time and with a minimum resolution
such that fluctuations cannot be seen (see figures of Sec. V).

In Sec. II we exactly solve the particular model, while in Secs. III and IV
we use this solution in order to obtain the coefficients of the master and
Langevin equations. Sec. V contains the numerical results.

\section{The exact solution of the model}

Let $h$ be the following Hamiltonian,

\begin{equation}
h=\Omega \left| \Omega \right\rangle \left\langle \Omega \right|
+\sum\limits_{n=1}^N\omega _n\left| \omega _n\right\rangle \left\langle
\omega _n\right| +\sum\limits_{n=1}^Ng_n\left( \left| \Omega \right\rangle
\left\langle \omega _n\right| +\left| \omega _n\right\rangle \left\langle
\Omega \right| \right) .  \label{fri}
\end{equation}
The eigenvalue problem, $h\left| \alpha _\nu \right\rangle =\alpha _\nu
\left| \alpha _\nu \right\rangle ,$ is reduced to the algebraic system

\begin{equation}
\begin{array}{c}
\Omega \left\langle \Omega |\alpha _\nu \right\rangle
+\sum\limits_{n=1}^Ng_n\left\langle \omega _n|\alpha _\nu \right\rangle
=\alpha _\nu \left\langle \Omega |\alpha _\nu \right\rangle , \\ 
\\ 
g_n\left\langle \Omega |\alpha _\nu \right\rangle +\omega _n\left\langle
\omega _n|\alpha _\nu \right\rangle =\alpha _\nu \left\langle \omega
_n|\alpha _\nu \right\rangle .
\end{array}
\label{sel}
\end{equation}
From the second of Eqs. (\ref{sel}) we have

\begin{equation}
\left\langle \omega _n|\alpha _\nu \right\rangle =\frac{g_n}{\alpha _\nu
-\omega _n}\left\langle \Omega |\alpha _\nu \right\rangle ,  \label{phi}
\end{equation}
for $\alpha _\nu \neq \omega _n,$ $\forall \nu ,n.$ Replacing Eq. (\ref{phi}%
) in the first equation of (\ref{sel}) we obtain the secular equation $%
\alpha _\nu -\Omega =\sum\limits_{n=1}^N\frac{g_n^2}{\alpha _\nu -\omega _n}%
. $ Finally, by using Eq. (\ref{phi}) and the completeness relation $\left|
\Omega \right\rangle \left\langle \Omega \right| +\sum\limits_{n=1}^N\left|
\omega _n\right\rangle \left\langle \omega _n\right| =I,$ we have

\begin{equation}
\left| \left\langle \Omega |\alpha _\nu \right\rangle \right| ^2=\frac 1{%
1+\sum\limits_{n=1}^N\left( \frac{g_n}{\alpha _\nu -\omega _n}\right) ^2},
\label{nor}
\end{equation}
where we have taken into account that $\left\langle \alpha _\nu |\alpha _\nu
\right\rangle =1.$

\section{The exact master equation}

As was showed in I the general solution of the master equation is given by ($%
n=0,...,N)$ $\left\langle N_n(t)\right\rangle =\sum_nP_{nm}(t)\left\langle
N_m(0)\right\rangle ,$ which in the particular model given by Eq. (\ref{fri}%
) reads

\begin{eqnarray}
\left\langle N_\Omega (t)\right\rangle &=&P_{\Omega \Omega }(t)\left\langle
N_\Omega (0)\right\rangle +\sum\limits_{n=1}^NP_{\Omega n}(t)\left\langle
N_n(0)\right\rangle ,  \nonumber \\
&&  \label{edgar} \\
\left\langle N_n(t)\right\rangle &=&P_{n\Omega }(t)\left\langle N_\Omega
(0)\right\rangle +\sum\limits_{m=1}^NP_{nm}(t)\left\langle
N_m(0)\right\rangle .  \nonumber
\end{eqnarray}
From Eq. (13) of I we have

\[
P_{nm}(t)=\sum_{\mu ,\nu =0}^Ne^{-i(\alpha _\mu -\alpha _\nu )t}\left\langle
\alpha _\mu |\psi _n\right\rangle \left\langle \alpha _\nu |\psi
_n\right\rangle \left\langle \psi _m|\alpha _\mu \right\rangle \left\langle
\psi _m|\alpha _\nu \right\rangle . 
\]
By taking into account the exact solution of the model [Eqs. (\ref{phi}) and
(\ref{nor})] these probabilities are given by

\begin{eqnarray}
P_{\Omega \Omega }(t) &=&2\sum\limits_{%
{\mu ,\nu =0 \atopwithdelims.. \mu >\nu }
}^N\frac{\cos \left[ \left( \alpha _\mu -\alpha _\nu \right) t\right] }{%
\left| \left\langle \Omega |\alpha _\mu \right\rangle \right| ^{-2}\left|
\left\langle \Omega |\alpha _\nu \right\rangle \right| ^{-2}}%
+\sum\limits_{\nu =0}^N\frac 1{\left| \left\langle \Omega |\alpha _\nu
\right\rangle \right| ^{-4}},  \nonumber \\
&&  \nonumber \\
P_{\Omega n}(t) &=&P_{n\Omega }(t)=2\sum\limits_{%
{\mu ,\nu =0 \atopwithdelims.. \mu >\nu }
}^N\frac{g_n^2\cos \left[ \left( \alpha _\mu -\alpha _\nu \right) t\right] }{%
\left| \left\langle \Omega |\alpha _\mu \right\rangle \right| ^{-2}\left|
\left\langle \Omega |\alpha _\nu \right\rangle \right| ^{-2}(\alpha _\mu
-\omega _n)(\alpha _\nu -\omega _n)}  \nonumber \\
&&\ +\sum\limits_{\nu =0}^N\frac{g_n^2}{\left| \left\langle \Omega |\alpha
_\nu \right\rangle \right| ^{-4}(\alpha _\nu -\omega _n)^2},  \label{pronm}
\\
P_{nm}(t) &=&2\sum\limits_{%
{\mu ,\nu =0 \atopwithdelims.. \mu >\nu }
}^N\frac{g_n^2g_m^2\cos \left[ \left( \alpha _\mu -\alpha _\nu \right)
t\right] }{\left| \left\langle \Omega |\alpha _\mu \right\rangle \right|
^{-2}\left| \left\langle \Omega |\alpha _\nu \right\rangle \right|
^{-2}(\alpha _\mu -\omega _n)(\alpha _\nu -\omega _n)(\alpha _\mu -\omega
_m)(\alpha _\nu -\omega _m)}  \nonumber \\
&&\ +\sum\limits_{\nu =0}^N\frac{g_n^2g_m^2}{\left| \left\langle \Omega
|\alpha _\nu \right\rangle \right| ^{-4}\left( \alpha _\nu -\omega _n\right)
^2\left( \alpha _\nu -\omega _m\right) ^2},  \nonumber
\end{eqnarray}
where $\left\{ \left| \psi _0\right\rangle ,\left| \psi _n\right\rangle
\right\} =\left\{ \left| \Omega \right\rangle ,\left| \omega _n\right\rangle
\right\} $ ($n=1,...,N).$

\section{The exact Langevin equation}

As it was shown in I the solution of the Langevin equation is

\begin{equation}
X(t)=\frac 12\left[ A_{\Omega \Omega }(t)+A_{\Omega \Omega }^{*}(t)\right]
X(0)+\frac 1{2i}\left[ A_{\Omega \Omega }(t)-A_{\Omega \Omega
}^{*}(t)\right] \frac{P(0)}M+f(t),  \label{le}
\end{equation}
which together with $A_{\Omega \Omega }(t)=\sum_{\nu =0}^N\frac{e^{-i\alpha
_\nu t}}{\left| \left\langle \Omega |\alpha _\nu \right\rangle \right| ^{-2}}
$ yields 
\[
X(t)=\sum_{\nu =0}^N\frac{\cos (\alpha _\nu t)}{\left| \left\langle \Omega
|\alpha _\nu \right\rangle \right| ^{-2}}X(0)+\sum_{\nu =0}^N\frac{\sin
(\alpha _\nu t)}{\left| \left\langle \Omega |\alpha _\nu \right\rangle
\right| ^{-2}}\frac{P(0)}M+f(t). 
\]
The coefficients result to be

\begin{equation}
\Omega ^2(t)=\frac{\sum\limits_{\mu ,\nu }\frac{\cos \left[ (\alpha _\nu
-\alpha _\mu )t\right] \alpha _\nu \alpha _\mu ^2}{\left| \left\langle
\Omega |\alpha _\mu \right\rangle \right| ^{-2}\left| \left\langle \Omega
|\alpha _\nu \right\rangle \right| ^{-2}}}{\sum\limits_{\mu ,\nu }\frac{\cos
\left[ (\alpha _\nu -\alpha _\mu )t\right] \alpha _\nu }{\left| \left\langle
\Omega |\alpha _\mu \right\rangle \right| ^{-2}\left| \left\langle \Omega
|\alpha _\nu \right\rangle \right| ^{-2}}},\hspace{0.3in}\Gamma (t)=-\frac{%
\sum\limits_{\mu ,\nu }\frac{\sin \left[ (\alpha _\nu -\alpha _\mu )t\right]
\alpha _\nu ^2}{\left| \left\langle \Omega |\alpha _\mu \right\rangle
\right| ^{-2}\left| \left\langle \Omega |\alpha _\nu \right\rangle \right|
^{-2}}}{\sum\limits_{\mu ,\nu }\frac{\cos \left[ (\alpha _\nu -\alpha _\mu
)t\right] \alpha _\nu }{\left| \left\langle \Omega |\alpha _\mu
\right\rangle \right| ^{-2}\left| \left\langle \Omega |\alpha _\nu
\right\rangle \right| ^{-2}}}.  \label{omega}
\end{equation}
These coefficients are independent of the initial condition which reflects
the purely dynamical evolution of the Brownian particle. The privileged
initial condition is only needed for computing the mean values of the
relevant observables leading to the apparently irreversible behavior from a
macroscopical point of view.

\section{Numerical results}

The parameters and functions that play the game are chosen as follows. The
mean values are taken with respect to the canonical distribution for the
bath, which is in thermal equilibrium at temperature $\beta ^{-1}$ at $t=0.$
The initial population for the Brownian oscillator is fixed to the unity and
its natural frequency $\Omega =\beta ^{-1}.$ The frequencies of the bath
oscillators are spaced with a constant step $A$ and centered around $\Omega $
according to $\omega _n=\Omega +A\left( n-\frac N2\right) ,$ where $A=\omega
_{n+1}-\omega _n;$ $n=1,...,N.$ The coupling function is a Lorentzian given
by $g_n=\frac{Aa^2}{a^2+\left( \omega _n-\Omega \right) ^2},$ where $a=\frac{%
A(N-2)}2.$ A more detailed discussion about this choice can be found in Ref. 
\cite{ggg}.

All the figures below correspond to $N=100,$ $A=0.018,$ $\Omega =1.$

\subsection{Master equation}

In Fig. 1 we have the time evolution of $\left\langle N_\Omega \right\rangle 
$ [given in Eq. (\ref{edgar})] departing from an initial population $%
\left\langle N_\Omega (0)\right\rangle =1.$ We can see that after a short
non-exponential regime (known as Zeno period \cite{zeno}) the decay profile
fits a decreasing exponential until it reaches the equilibrium value $%
\left\langle N_\Omega (t\gg \Omega ^{-1},\gamma ^{-1})\right\rangle =\left(
e^{\beta \Omega }-1\right) ^{-1}\cong 0.58$ (where the Brownian oscillator
thermalizes with the bath). It is the asymptotic value reached by $%
\left\langle N_\Omega \right\rangle $ for times smaller than the recurrence
time $t_r\approx \frac{2\pi }{{\rm \min }(\alpha _{\nu +1}-\alpha _\nu )}%
\cong 37,311,$ which is several orders of magnitude greater than the
oscillator period $\tau _\Omega =2\pi /\Omega =2\pi $ (see a detailed
explanation in Ref. \cite{ggg}). Fig. 2 shows the contributions to $%
\left\langle N_\Omega \right\rangle $ stemming from the survival probability 
$P_{\Omega \Omega }(t)=\left| \left\langle \Omega \left| e^{-iht}\right|
\Omega \right\rangle \right| ^2$ of the initially prepared unstable state $%
\left| \Omega \right\rangle ,$ which reaches for $t<t_r$ a vanishing
asymptotic value, and the contribution coming from the bath $\sum_nP_{\Omega
n}(t)\left\langle N_n(0)\right\rangle ,$ which provides the equilibrium
value to $\left\langle N_\Omega \right\rangle .$

\subsection{Langevin equation}

In Fig. 3 we show the damped oscillations of the Brownian particle comparing
it with the reconstruction and decay of $\left\langle N_\Omega \right\rangle 
$ in the peak around the recurrence time. Fig. 4 gives the time behavior of
the $`$`damping'' coefficient $\Gamma (t)$ re-scaled with respect to its
asymptotic value $\gamma $ obtained in I through second order perturbation
theory. It can be shown that $\Gamma (t)$ essentially follows $-2{\rm Re}%
\frac{\stackrel{.}{A_{\Omega \Omega }}}{A_{\Omega \Omega }}$ at all times.
We firstly see that $\Gamma (t)$ grows from 0 to $\gamma $, corresponding to
the Zeno regime. Secondly, it oscillates around $\Gamma =\gamma $ as a
consequence of the presence of fluctuations which modulates the exponential
decay. After that, the amplitude of these oscillations increases in time
because of the relative variations between fluctuations and the exponential
decay ($-2{\rm Re}\frac{\stackrel{.}{A_{\Omega \Omega }}}{A_{\Omega \Omega }}%
)$ become more important when the system reaches equilibrium. Finally, we
see the effect of recurrences also in $\Gamma (t).$ It is interesting to
stress the fact that $\Gamma (t)$ allows us to visualize the presence of
fluctuations which, at the same time scale, are hidden in the observable
macroscopic profiles.

\newpage\ 

\noindent
{\bf Figure captions:}

1. Population of the Brownian oscillator {\it vs.} $t$

2. Survival probability {\it vs.} $t$

3. Mean position of the Brownian oscillator {\it vs.} $t$

4. Damping factor of the Langevin equation {\it vs.} $t$

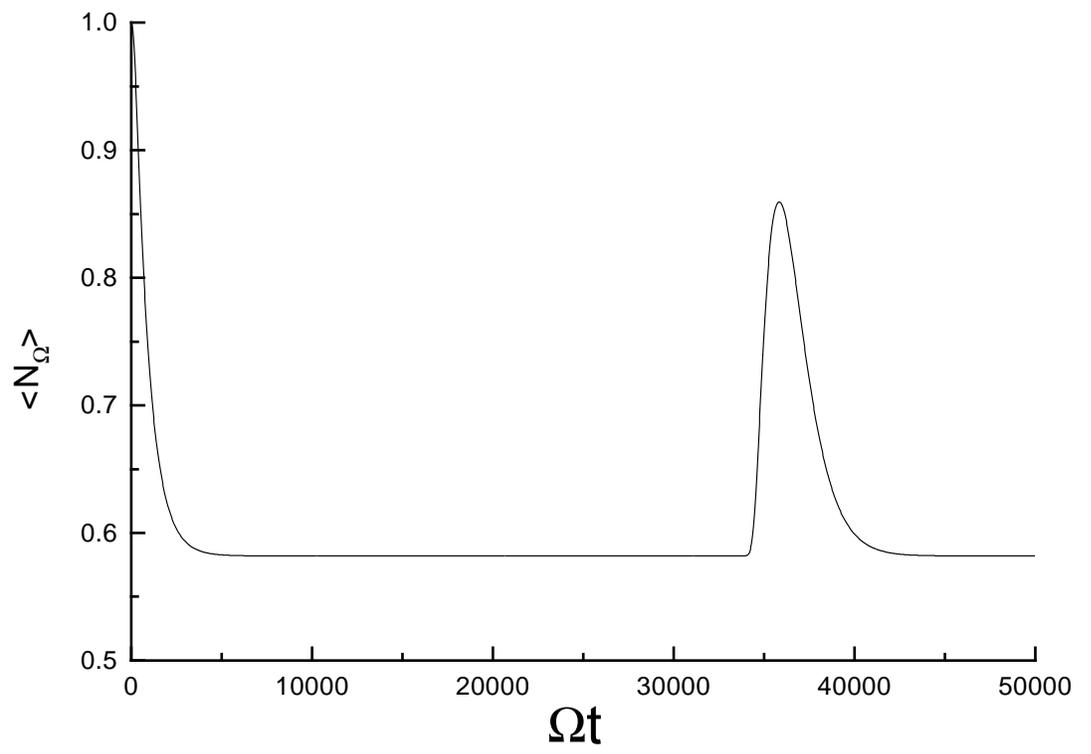
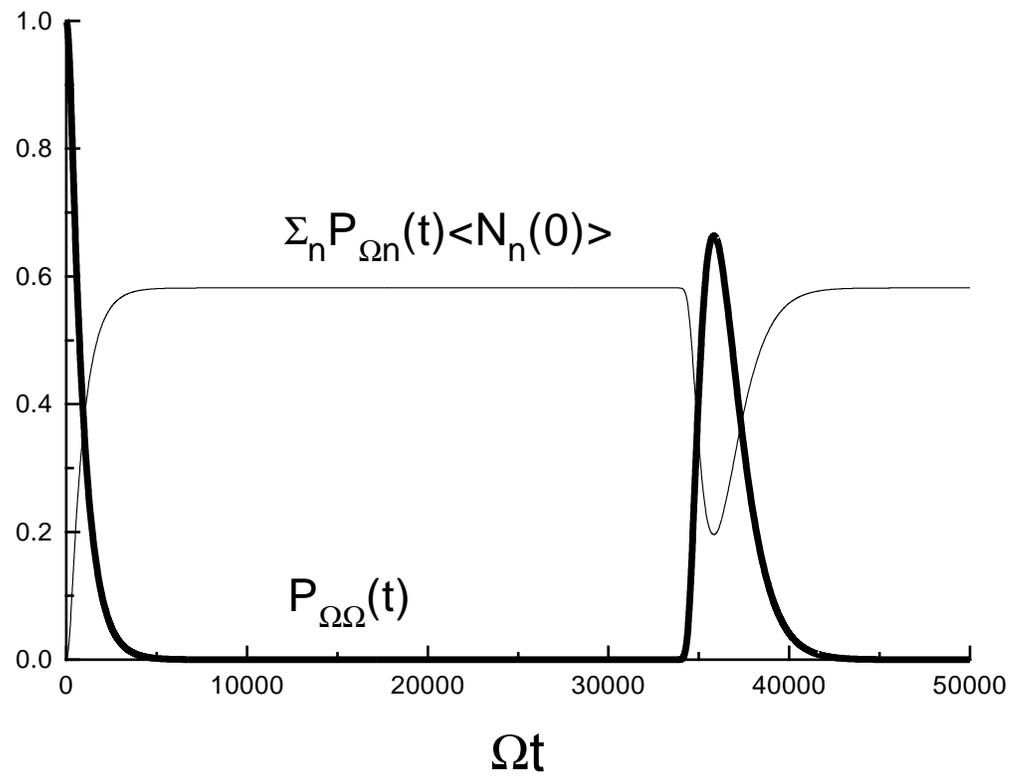
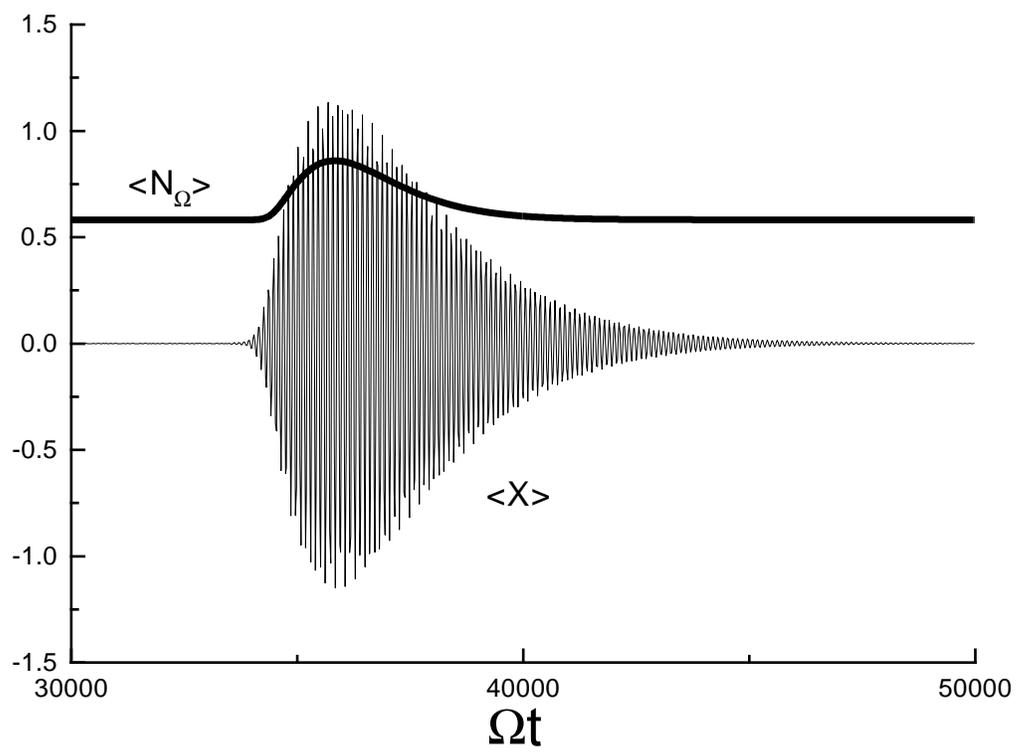
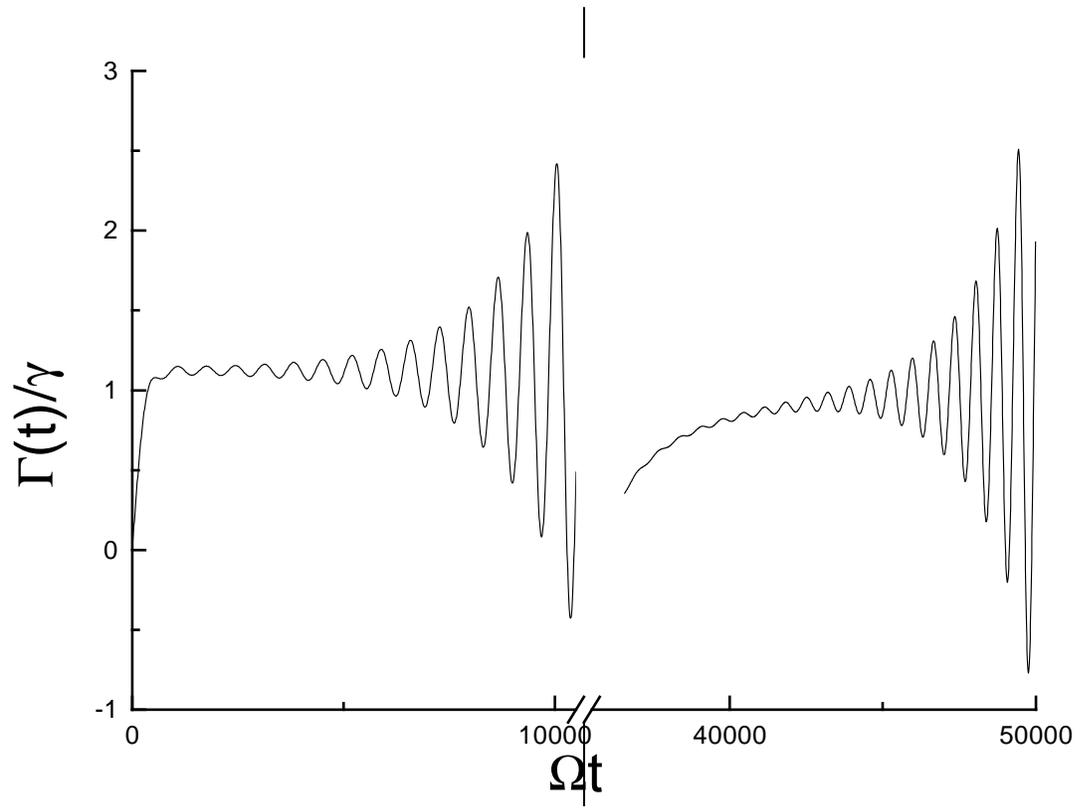
\end{document}